\documentclass[10pt,aps,twocolumn,floatfix,showpacs,showkeys,preprintnumbers,nofootinbib,superscriptaddress,longbibliography,bibnotes]{revtex4-1}
\pdfoutput=1
\usepackage[colorlinks=true,breaklinks=true]{hyperref}
\usepackage[utf8]{inputenc}
\hypersetup{allcolors=[rgb]{0.0 0.0 0.6},linkcolor=[rgb]{0.75 0.05 0.05}}
\usepackage{amsmath,amssymb}
\usepackage{epsfig}  
\usepackage{graphicx}   
\usepackage{slashed}             
\usepackage{url}
\usepackage{color}
\usepackage{multirow}
\usepackage{placeins}
\usepackage[dvipsnames]{xcolor}
\usepackage{letltxmacro}
\LetLtxMacro{\oldcite}{\cite}
\renewcommand{\cite}[1]{\mbox{\oldcite{#1}}}
\usepackage{enumitem}

\usepackage[normalem]{ulem}

\newcommand{\G}{G}

\allowdisplaybreaks

\bibpunct{[}{]}{,}{n}{}{,}

\makeatletter
\g@addto@macro\bfseries{\boldmath}
\makeatother

\begin{document}

\preprint{MPP-2020-224}

\title{On the characteristics of fast neutrino flavor instabilities in three-dimensional core-collapse supernova models} 

\author{Sajad Abbar}
\email{abbar@mppmu.mpg.de}
\affiliation{Max Planck Institut f\"{u}r Physik (Werner-Heisenberg-Institut), F\"{o}hringer Ring 6, 80805 M\"{u}nchen, Germany. 
}                       
\author{Francesco Capozzi}
\email{capozzi@mppmu.mpg.de}
\affiliation{Max Planck Institut f\"{u}r Physik (Werner-Heisenberg-Institut), F\"{o}hringer Ring 6, 80805 M\"{u}nchen, Germany. 
} 
\affiliation{Center for Neutrino Physics, Department of Physics, Virginia Tech, Blacksburg, VA 24061, USA}
\author{Robert~Glas}
\email{rglas@mpa-garching.mpg.de}
\affiliation{Max-Planck-Institut f\"ur Astrophysik, Karl-Schwarzschild-Stra{\ss}e 1, D-85748 Garching, Germany}
\affiliation{Excellence Cluster ORIGINS, Boltzmannstr.~2, D-85748 Garching, Germany}             
\author{H.-Thomas~Janka}
\email{thj@mpa-garching.mpg.de}
\affiliation{Max-Planck-Institut f\"ur Astrophysik, Karl-Schwarzschild-Stra{\ss}e 1, D-85748 Garching, Germany}
\author{Irene Tamborra}
\email{tamborra@nbi.ku.dk}
\affiliation{Niels Bohr International Academy and DARK, Niels Bohr Institute, University of Copenhagen, Blegdamsvej 17, 2100, Copenhagen, Denmark. }


\begin{abstract}
We assess the occurrence of fast neutrino flavor instabilities
 in two three-dimensional state-of-the-art core-collapse supernova simulations performed using a two-moment three-species neutrino transport scheme: one with an exploding 9$\mathrm{M_{\odot}}$ and one with a non-exploding  20$\mathrm{M_{\odot}}$ model.
Apart from confirming the presence of fast instabilities occurring within the neutrino decoupling and  the supernova pre-shock regions,
we detect 
flavor instabilities in the post-shock region for the exploding model. These instabilities are likely
 to be scattering-induced.
In addition, the failure in achieving a successful explosion in the heavier supernova model seems to seriously 
hinder the occurrence of fast instabilities in the post-shock region.
This is a consequence of the large matter densities behind the stalled or retreating shock, which implies high neutrino scattering
 rates and thus more isotropic distributions of neutrinos and antineutrinos. 
 Our findings suggest that the supernova model properties and the fate of the explosion can remarkably affect 
 the occurrence of fast instabilities. Hence, a larger set of realistic hydrodynamical simulations of the stellar collapse 
 is needed in order to make reliable predictions on the flavor conversion physics. 
\end{abstract}

\maketitle

\section{Introduction}

Massive stars can end their lives in the form of core-collapse supernova (CCSN)
explosions, where a huge amount of energy ($\sim 3\times10^{53}$~ergs) is released  of which 
 almost $99\%$ is in the form of neutrinos
 of all flavors~\cite{Janka:2012wk, Burrows:2012ew, Mezzacappa:2005ju}. 
In this context, neutrino flavor evolution can be a collective phenomenon
  in which neutrinos propagating with different momenta change their flavor altogether thanks
 to the neutrino-neutrino interactions in the SN environment~\cite{Pastor:2002we,duan:2006an, duan:2010bg}.

 Neutrino flavor conversions are important to the physics of CCSNe. In particular,
they can change the neutrino and antineutrino energy distributions and consequently
 their scattering rates. This can, at least in principle, impact the CCSN dynamics
and the nucleosynthesis of heavy elements in the SN environment~\cite{Qian:1996xt}. Moreover,
 understanding collective neutrino oscillations is necessary for analyzing the neutrino signal
  from the next galactic CCSN and the upcoming measurements of the diffuse SN neutrino
  background~\cite{Mirizzi:2015eza, Scholberg:2012id, Lunardini:2010ab, Beacom:2010kk}.

It has been shown that neutrinos can  experience
\emph{fast} flavor conversions in  dense neutrino 
media~\cite{Sawyer:2005jk, Sawyer:2015dsa,
 Chakraborty:2016lct, Izaguirre:2016gsx,  Wu:2017qpc,
  Capozzi:2017gqd,  
   Abbar:2017pkh, Abbar:2018beu, Capozzi:2018clo, Yi:2019hrp,
 Martin:2019gxb, Capozzi:2019lso,  Johns:2019izj,
 Shalgar:2019qwg, Cherry:2019vkv, Chakraborty:2019wxe,  Abbar:2020fcl, Capozzi:2020kge, Xiong:2020ntn, Bhattacharyya:2020dhu,
 Shalgar:2020xns, Johns:2020qsk, Bhattacharyya:2020jpj, George:2020veu, 
 Shalgar:2020wcx, Tamborra:2020cul}.
 In contrast to the
traditional slow modes which occur on scales determined by the vacuum frequency
 ($\gtrsim \mathcal{O}(1)$~km for a typical SN neutrino), fast modes occur
 on scales $\sim G_{\mathrm{F}}^{-1} n_{\nu}^{-1}$, where
 $G_{\mathrm{F}} $ is the Fermi constant and $ n_{\nu}$ is the neutrino number density.
 Such scales can be
  as short as a few cm 
  in the vicinity of the proto-neutron star (PNS).
  Although both fast and slow modes can be triggered below the shock (for the latter this has been 
 recently shown to be possible through matter turbulence \cite{Abbar:2020ror}),
  fast modes remain unique because in the deepest regions of the SN core they can  develop on scales $\sim$ few cm, 
  which are much shorter than the average neutrino scattering length.

It is widely believed that a necessary (and probably sufficient) condition 
for the occurrence of fast flavor instabilities is that the angular distributions of $\nu_e$ and
$\bar\nu_e$ cross each 
other~\cite{Izaguirre:2016gsx,Dasgupta:2016dbv,Capozzi:2017gqd, Abbar:2017pkh,Yi:2019hrp,Capozzi:2019lso}.
To be more precise, the angular distribution of the neutrino electron lepton number (ELN) should change its sign at a specific angle, i.e. it should have a crossing (see Sec.~\ref{sec:method} for
details).
The first assessments of the occurrence of ELN crossings in CCSNe
were based on the analytical 
two-bulb model~\cite{Sawyer:2005jk,Sawyer:2008zs,Sawyer:2015dsa,Chakraborty:2016yeg} in which two separate neutrinospheres were considered for $\nu_e$ and
$\bar\nu_e$. 
Although  ELN crossings are always present 
  in such models, at least as long 
as the neutrino-antineutrino asymmetry parameter 
\begin{equation}
\alpha = n_{\bar\nu_e}/n_{\nu_e},
\end{equation}
 is not very small,
it was soon realized that their existence in realistic CCSN simulations is not  guaranteed. 
In particular, ELN crossings were not found in the post-shock region of one-dimensional (1D)
CCSN simulations~\cite{Tamborra:2017ubu,Morinaga:2019wsv}. However,  recently several studies have reported the existence
of ELN crossings and their associated fast instabilities in multidimensional  (multi-D)  simulations~\cite{Abbar:2018shq,DelfanAzari:2019tez,Nagakura:2019sig,Abbar:2019zoq,Glas:2019ijo}.
Fast instabilities can indeed occur 
via different mechanisms in  different  SN regions:

\begin{itemize}[leftmargin=*]
 \item
\textbf{In the convective layer of the PNS}, well below the neutrinospheres  
where the angular distributions of $\nu_e$ and $\bar\nu_e$ are still almost isotropic~\cite{DelfanAzari:2019tez, Abbar:2019zoq,Glas:2019ijo}.
Such a possibility originates from the strong convective activity therein, which can generate  
large amplitude modulations in the spatial distributions of $\nu_e$ and $\bar\nu_e$
number densities.  Hence,  zones  can exist   there,
for which  $\alpha$ is   arbitrarily close to one and no matter 
how isotropic are the distributions of $\nu_e$ and $\bar\nu_e$, ELN crossings can always
 occur. However and despite such an interesting possibility, the physical implications of this class of ELN
 crossings are not very clear due to the
  nearly equal distributions of neutrinos and antineutrinos of all flavors (with chemical potential close to zero), which
 means that neutrino oscillations cannot significantly  modify
 the flavor content of the neutrino gas.

\item \textbf{Within the neutrino decoupling region
 just above the PNS}~\cite{Abbar:2018shq,Nagakura:2019sig,Abbar:2019zoq}.  The existence of ELN crossings in this region can be explained 
 as follows. Due to the neutron richness of the SN matter, $\nu_e$'s decouple at
 larger radii than $\bar\nu_e$'s. As a consequence, the  angular distributions of $\bar\nu_e$'s
 are in general more peaked in the forward direction than those of $\nu_e$'s, thus increasing 
 the chance for the occurrence of ELN crossings. Nevertheless,  
 ELN crossings can still only occur in zones  where $\alpha$ is relatively close to one and
 are suppressed otherwise~\cite{Abbar:2018shq, Shalgar:2019kzy}. 
 This means that asymmetry in the neutrino emission,
  caused by  the LESA (lepton-emission self-sustained asymmetry) 
  phenomenon~\cite{Tamborra:2014aua, Glas:2018vcs}, 
  plays an important role here (other effects leading to  asymmetric neutrino emission were discussed, e.g., in Ref.~\cite{Nagakura:2019evv}).
 Indeed the associated asymmetric  lepton number emission implies that the distribution of $\alpha$
 is not uniform above the PNS and there can exist zones for which $\alpha$
 is relatively close to one. 
 This increases  the chance for the occurrence of  ELN crossings in such SN zones.
 Needless to say, this type of ELN crossings 
 should be then seriously suppressed if LESA is not strong enough.

\item \textbf{In the pre-shock 
SN region}. As first pointed out in Ref.~\cite{Morinaga:2019wsv}, this type of ELN crossings is generated
 by neutrino backward scatterings off heavy nuclei. 
  At a fixed energy, the scattering rate of neutrinos is larger than the one of antineutrinos. 
However, taking into account that the latter have a larger average energy, 
one finds that $\bar{\nu}_e$ has a larger cross section. 
This feature allows 
antineutrinos to be dominant in the backward direction
and generate a backward ELN crossing 
 despite the fact that neutrinos have an overall larger number density. 
 This sort of ELN
 crossings can be more generic than
 the  ELN crossings discussed previously. In spite of this, 
 such fast instabilities  cannot impact the SN (shock) dynamics  and in addition, 
 their corresponding growth rates could be very small  
 due to the smaller $n_\nu$  in the SN pre-shock region. 
 
\end{itemize}   

Considering the discussion above, perhaps the fast instabilities  occurring within the
neutrino decoupling region have the most important physical implications. Here, a remark is in order.
In almost all the state-of-the-art 3D CCSN simulations the equations for the neutrino transport are solved in terms of a few angular moments
 of the neutrino phase space distribution. 
 This is done to avoid the extremely demanding computational requirements of solving the Boltzmann equation. 
 Such an approximation prevents us from having direct information on the underlying neutrino angular distributions. 
 The only exception is the Ref.~\cite{Iwakami:2020ctd} in 
 which a 3D CCSN simulation was performed with full
Boltzmann neutrino transport,
 but only up to 20 ms after bounce  and still with severe limitations of the numerical resolution
 (see, e.g. Refs.~\cite{Nagakura:2017mnp, Nagakura:2019evv, Harada:2018ubo, Harada:2020fek, Chan:2020quo} for examples of CCSN simulations with full
Boltzmann neutrino transport). 
  Most of the previous studies of the occurrence of fast instabilities  have been  based on either 1D/2D SN models or post-processing calculations
   in which the full neutrino Boltzmann equation is solved for static  SN density profiles derived from less fashionable CCSN simulations. 
   One may wonder whether the results obtained in this context can be confirmed/improved
   when  the state-of-the-art 3D CCSN simulations are considered. 
   In fact the only study adopting such models found the occurrence of fast instabilities only in the convective layer of the PNS~\cite{Glas:2019ijo}.

Although a significant amount of information gets lost using the neutrino angular moments, 
 it is still possible to extract some useful information regarding the occurrence of ELN crossings.  
 In particular, it has been shown that ELN crossings are guaranteed to exist if 
 the neutrino angular moments meet certain criteria, as discussed in Sec.~\ref{sec:method}. 
 In this paper, we assess the possibility of the occurrence of fast instabilities in  the SN 
 environment by studying
two 3D CCSN simulations of a 9$\mathrm{M}_{\odot}$ and  a $20\mathrm{M}_\odot$  model,  which
were recently performed by the Garching group~\cite{Glas:2018vcs,Glas:2018oyz} 
(discussed briefly in Sec.~\ref{sec:SN}). 
As we discuss in Sec.~\ref{sec:results},
not only do we confirm the possibility of the occurrence of fast instabilities within the neutrino decoupling 
and the pre-shock SN regions, 
but also we bring up a new possibility for the occurrence of  probably scattering-induced fast instabilities
in the post-shock SN region.
Moreover, we demonstrate that there can be an important connection between
the fate of the explosion and the chance for the occurrence of fast instabilities in CCSNe,
 i.e.
the occurrence of ELN crossings can be seriously hindered in non-exploding SN models with high mass accretion rates.

\section{supernova models}\label{sec:SN}

Our study 
is based on two self-consistent 3D CCSN simulations
 carried out by the Garching group~\cite{Glas:2018oyz,Glas:2018vcs}:
in the first simulation, a 9M$_{\odot}$  model successfully exploded
via the delayed neutrino-heating mechanism with the shock revival setting in at about 300\,ms after the core bounce,
whereas in the second one, a more massive progenitor star with 20M$_{\odot}$ failed to explode.
Both simulations were performed with the \textsc{Aenus-Alcar} code~\cite{Obergaulinger:2008pb, Just:2015fda,Just:2018djz},
which solves the hydrodynamics equations describing the time-dependent flow properties of the stellar plasma,
and the equations of a velocity- and energy-dependent two-moment neutrino transport scheme.
In this so-called ``M1'' scheme, the computationally expensive solution of the Boltzmann transport equation  is circumvented by 
the approximation of evolving the first two angular moment equations of the Boltzmann transport equation. This implies that the transport solver does not
compute the time-dependent neutrino phase-space distribution functions, but instead it tracks the evolution of the energy-dependent
moments of neutrino energy density and neutrino energy-flux density.

While in the original paper~\cite{Glas:2018oyz} the authors varied both the neutrino transport approximation and the numerical grid resolution,
we consider here only the fiducial models with fully multi-D (FMD) neutrino transport and with higher grid resolution,
i.e., models ``s9.0~FMD~H'' and ``s20~FMD~H'' for the 9M$_{\odot}$ and 20M$_{\odot}$ progenitor stars, respectively.
These models were simulated in spherical polar coordinates
$r$ (radius), $\Theta$ (polar angle), and $\Phi$ (azimuthal angle)
 with an effective angular resolution of roughly $2^\circ$,
and with 640 radial zones spanning a range from $0$ to $10^4\,\mathrm{km}$.

In both simulations, the neutrino transport equations were solved for electron neutrinos ($\nu_e$),
electron anti-neutrinos ($\bar{\nu}_e$), and a third species $\nu_x$ that represents all four kinds of heavy-lepton neutrinos
($\nu_\mu$, $\bar{\nu}_\mu$, $\nu_\tau$, and $\bar{\nu}_\tau$), whose transport properties are considered to be identical.
The interactions between neutrinos and the stellar matter included a whole set of absorption, emission, scattering,
and thermal production processes, for which we refer the reader to the original publication.

\section{Searching for fast instabilities}\label{sec:method}
As mentioned above, 
a necessary condition for the existence of fast flavor instabilities in a dense neutrino gas is the presence of crossings in the ELN angular distribution. The latter is defined as
\begin{equation}
  \G_\mathbf{v} =
  \sqrt2 G_{\mathrm{F}}
  \int_0^\infty \frac{E_\nu^2 \mathrm{d} E_\nu}{(2\pi)^3}
        [f_{\nu_e}(\mathbf{p})- f_{\bar\nu_e}(\mathbf{p})],
\end{equation}
where $E_\nu$, $\mathbf{p}$, and $\mathbf{v}$
are the neutrino energy, momentum, and velocity, respectively,
and $f_\nu$'s are the neutrino occupation numbers. Here, $G_{\mathrm{F}}$ is the Fermi constant.
Note that this definition is useful only when  $f_{\nu_{\mu,\tau}}=f_{\bar{\nu}_{\mu,\tau}}$.
In what follows, 
 we consider the $\phi_\nu$-integrated distribution of ELN
\begin{equation}
  \G(\mu) =
  \int_0^{2\pi}  \mathrm{d} \phi_\nu  \G_\mathbf{v},
\end{equation}
where $\mu=\cos\theta_\nu$, and $\theta_\nu$ and $\phi_\nu$ are  the  zenith and  azimuthal angles 
 of the neutrino momentum.
 Note that the integration over $\phi_\nu$ will not generate  any fake 
 crossing, though it might erase some of the ELN crossings occurring in $\phi_\nu$.
 We say that $G(\mu)$ has a crossing if it changes its sign at some values of $\mu$. 

In the M1 closure scheme, instead of the detailed ELN angular distributions one has access
to a few of its moments, defined as
 \begin{equation}
I_n = \int_{-1}^{1} \mathrm{d}\mu\ \mu^n\ G(\mu).
\label{moments}
\end{equation}
To be specific, apart from the evolutions of $I_0$ and $I_1$ which are computed numerically,
one has also access to $I_2$ and $I_3$ through analytical closure relations. 
Even with such a limited amount of information, the presence of ELN crossing(s) 
 can still be efficiently evaluated  as discussed recently in Ref.~\cite{Abbar:2020fcl}.
Here it was shown that ELN crossing(s) are guaranteed to occur at a given SN zone if there exists a  
positive function $\mathcal{F}(\mu)$, for which 
\begin{equation}
I_{\mathcal{F}}I_0<0,
\label{eq:crossing_condition}
\end{equation}
where
\begin{equation}
I_\mathcal{F} = \int_{-1}^{1} \mathrm{d}\mu\ \mathcal{F}(\mu)\  G(\mu).
\label{IF}
\end{equation}
Because the neutrino angular information is provided in terms of $I_n$'s,
we choose $\mathcal{F}(\mu)$ to be a polynomial of $\mu$
 \begin{equation}
 \label{eq:F}
\mathcal{F}(\mu) =\sum_{n=0}^N a_n \mu^n,
\end{equation}
where $N$ is the highest  neutrino angular moment available from numerical simulations
and $a_n$'s are arbitrary coefficients for which $\mathcal{F}(\mu)>0$.
$I_\mathcal{F}$  can then be written in terms 
of $I_n$'s
\begin{equation}
I_\mathcal{F}=a_0I_0+a_1I_1+\dots+ a_NI_N\,.
\label{eq:I_F}
\end{equation}

In order to assess the presence of fast instabilities in  realistic SN simulation data, for a number of representative post bounce time snapshots and each SN point, identified by the coordinates $(r,\Theta,\Phi)$, we do the following:
\begin{enumerate}
\item We extract the moments of the ELN up to $I_3$, where $I_2$ and $I_3$ are obtained from   
 $I_0$ and $I_1$ via Minerbo analytical closure relations (see Eqs.~(28a) and (33) in Ref.~\cite{Just:2015fda} and Ref.~\cite{minerbo1978maximum}). 
\item We perform a scan of all appropriate quadratic and cubic polynomial functional forms of $\mathcal F(\mu)$, i.e. for $a_3=0$ and $a_3\neq 0$
. If we find a polynomial such that
 Eq.~(\ref{eq:crossing_condition}) is satisfied, we mark the SN point and  we go back to step 1, 
 choosing a different point. 
\item If not, we apply a functional minimization routine to the product $I_0 I_{\mathcal F} $.
\item We repeat the previous steps around the minimum if it is negative.
\end{enumerate}

The search done with a cubic polynomial is intrinsically more powerful because it 
can capture  finer angular  structures and, therefore, narrower ELN crossings.
Nevertheless, we find that most of the crossings can be found also with a quadratic 
polynomial. This is  important since it allows us to do not completely rely on $I_3$ which 
is not as accurate as the lower moments~\cite{Just:2015fda}.

The CCSN data we are about to study were also recently investigated in 
Ref.~\cite{Glas:2019ijo}  where only a number of ELN crossings were found in the
convective layer inside the PNS. In that study, the  method proposed in 
Ref.~ \cite{Dasgupta:2018ulw}
was used, which is based on the instability of a particular Fourier mode (the so-called zeroth mode).
However as discussed in the next section,
in the same  CCSN simulations we find a much larger number of ELN crossings in the SN region
outside the PNS.
We have also checked that our method can capture 
 the same ELN crossings inside the PNS obtained in Ref.~\cite{Glas:2019ijo}.
 This is a confirmation of the power of our method, which can also provide a unique insight on the type of the ELN crossing based on the shape of the appropriate $\mathcal F(\mu)$'s.
 
 
As a final remark, we emphasize that our method might not capture all the ELN crossings.
Indeed, the narrower they are the higher the order of the moments required, i.e. $N>3$.
For example, by comparing the number of crossings seen directly with angular distributions
and the one found using the moments method with $N=3$, in Ref.~\cite{Abbar:2020fcl} it was estimated that  almost 50\% of crossings are 
missed when  the moments method is employed. However, in Ref.~\cite{Abbar:2020fcl} the moments are calculated directly
from the angular distributions, so such an estimate might not be directly applicable to our case.

\section{Fast instabilities in 9$\mathrm{M_{\odot}}$ and 20$\mathrm{M_{\odot}}$  models}\label{sec:results}

In this section, we present the results of our ELN crossing search  for a number of representative
 time snapshots, namely $t=100$, 200, 300, 400 and 500 ms after core bounce for each  model.
 For the sake of plotting, in this section we consider the ELN crossings captured when $a_3 \neq 0$ though
 as mentioned previously, the results do not change notably when  $a_3 = 0$.
  Furthermore,
  in what follows  we focus  on the ELN crossings in the region outside the PNS  to avoid any repetition
  of the results already discussed in Ref.~\cite{Glas:2019ijo}.

\subsection{$9\mathrm{M_{\odot}}$ supernova model}
\subsubsection*{$\mathbf{t=100}$  \rm{\textbf{ms}} }
At this time, the SN has not exploded yet and it is in the accretion phase
where the radius of the $\nu_e$ neutrinosphere is  $\sim50$ km and the shock radius is  $\sim150$ km.
We find ELN crossings starting from $r=91$ km (for which $\sqrt2 G_{\mathrm{F}} n_{\nu_e} = 6.9\times 10^4$ km$^{-1}$)
 up to distances much larger than the shock radius. 

\begin{figure}
  \centering
    \includegraphics[width=0.4\textwidth,  trim= 0 0 0 0,clip]{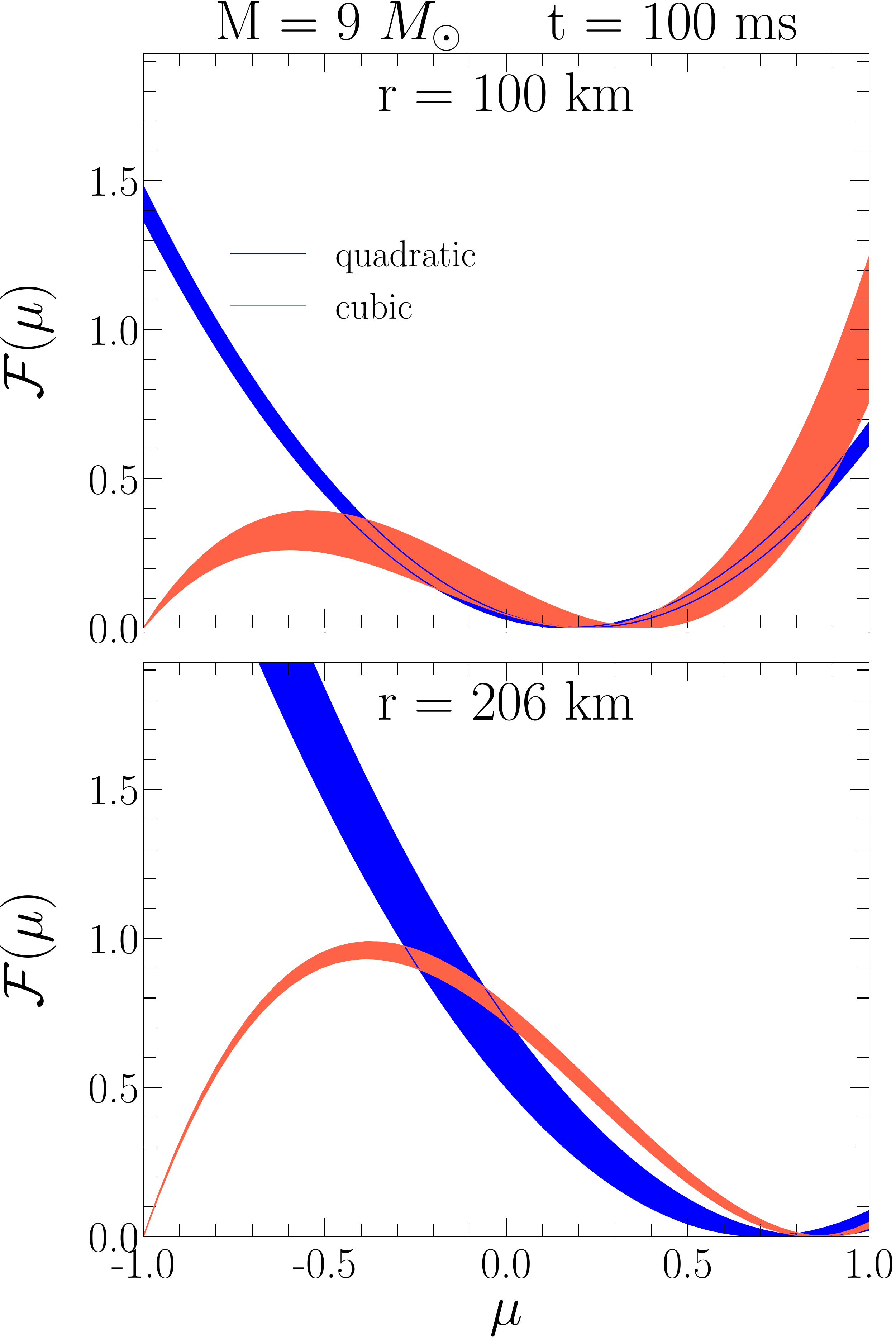}
\caption{Shapes of all $\mathcal{F}(\mu)$'s (at different spatial $\Theta$'s and $\Phi$'s) that satisfy Eq.~(\ref{eq:crossing_condition}) for $9\mathrm{M_{\odot}}$ model at $t = 100$ ms, and at $r = 100$~km (below the shock)
and $206$~km (above the shock), respectively.  
Red and  blue bands represent cubic and quadratic polynomial functions, respectively. 
The nature of the crossings changes around the shock, which is at $r \simeq150$ km.}
        \label{Fig:9Msol_100ms_F}
\end{figure}

In Fig.~\ref{Fig:9Msol_100ms_F}, we show the shapes of $\mathcal{F}(\mu)$ which allow for capturing the ELN  crossings, at two representative radial distances.
 At $r=100$ km (top panel), $\mathcal{F}(\mu)$ takes  large values for $\mu\to 1$, which implies that the actual ELN angular distributions are most likely
 negative there and the crossings are in the forward direction~\footnote{In order to capture the ELN crossings, $\mathcal{F}(\mu)$ should be relatively large at $\mu$'s
  where the sign of $G(\mu)$ is opposite to the sign of $I_0$~\cite{Abbar:2020fcl}.}.
 These are indeed the crossings that one expects within/above the neutrino decoupling region.
 However, this changes at larger radii ($r=206$ km, bottom panel) and specifically above the shock.
 In the pre-shock region, the ELN crossings are expected to be in the backward direction as
 discussed before. Hence, $\mathcal F(\mu)$ must be large in the backward direction to be capable of capturing  the  crossings.

\begin{figure}
  \centering
\includegraphics[width=.4\textwidth]{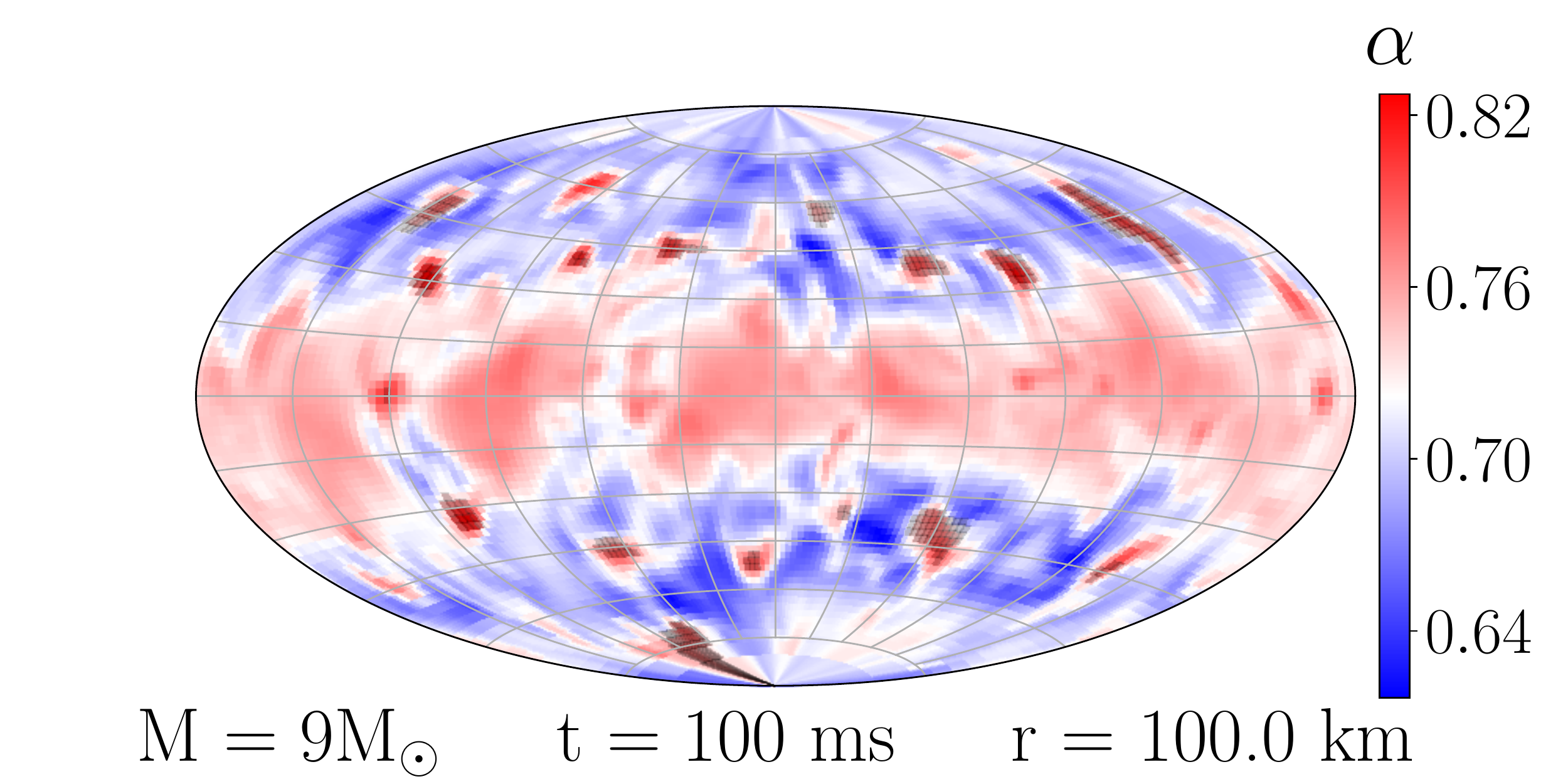}
  \caption{Aitoff projection of $\alpha$ in the $t=100$ ms snapshot at $r=100$ km. 
  The shaded zones  indicate the angular coordinates 
  for which we find ELN crossings in the range $[90,150]$ km. 
  Note that the crossings occur where $\alpha$ is maximal. This is a hint in favor of a correlation with the asymmetric PNS convection.}
 \label{Fig:9Msol_100ms_alpha}
\end{figure}

The spatial distributions of  the ELN crossings within/above the neutrino decoupling region are 
strongly correlated with the maxima of $\alpha$, as clearly shown in its 
Aitoff projection at $r=100$ km reported in Fig.~\ref{Fig:9Msol_100ms_alpha}.
Here, for the sake of presentation, we have shown all  the ELN crossings captured
between the $\nu_e$ neutrinosphere  and the shock. We also noted that the distribution of $\alpha$
doest not change noticeably in this range.
However, this changes for the backward crossings above the shock ($r>150$ km)
which are basically ubiquitous~\cite{Morinaga:2019wsv} (not presented here).

\subsubsection*{$\mathbf{t=200}$  \rm{\textbf{ms}} }
The model has not exploded yet and the radius of the $\nu_e$ neutrinosphere is  $\sim40$ km and the shock radius is  $\sim200$ km.
In this time snapshot, we find ELN crossings  starting from $r=136$ km (for which $\sqrt2 G_{\mathrm{F}} n_{\nu_e} = 1.2\times 10^4$ km$^{-1}$)
 up to distances much larger than the shock radius. 
Crossings below the shock are again probably in the forward direction (though assessment is not very easy in this case) and they occur in a very specific spatial direction where $\alpha$ is maximal. 
In addition, the spatial distribution of the backward ELN crossings is ubiquitous  above the shock,  as expected.
\subsubsection*{$\mathbf{t=300}$  \rm{\textbf{ms}} }
We find no crossings below the shock wave ($r<260$ km), 
whereas the ubiquitous backward crossings above the shock start from about 300 km. 
The absence of the forward ELN crossings in this time snapshot might be understood by
noting that as time goes by, 
the existing ELN crossings in the neutrino decoupling region gets narrower.
This means that one should have access to 
   higher neutrino angular moments to be able to capture the crossings.

\begin{figure*}
  \centering
    \includegraphics[width=\textwidth]{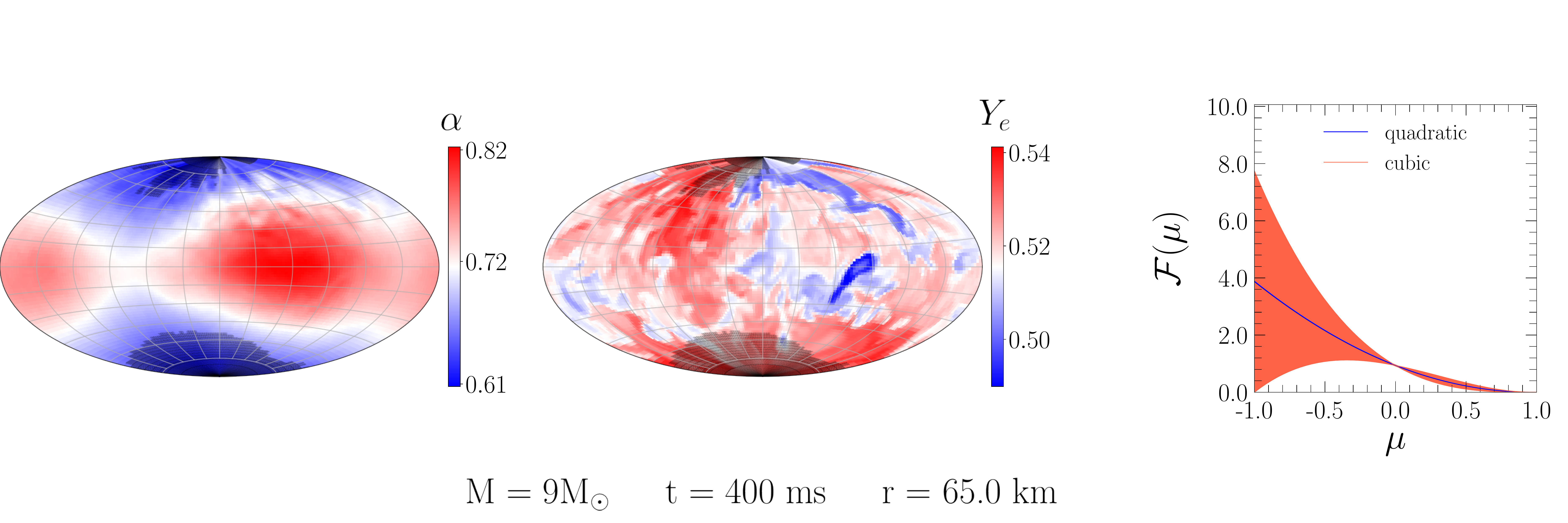}
          \includegraphics[width=\textwidth]{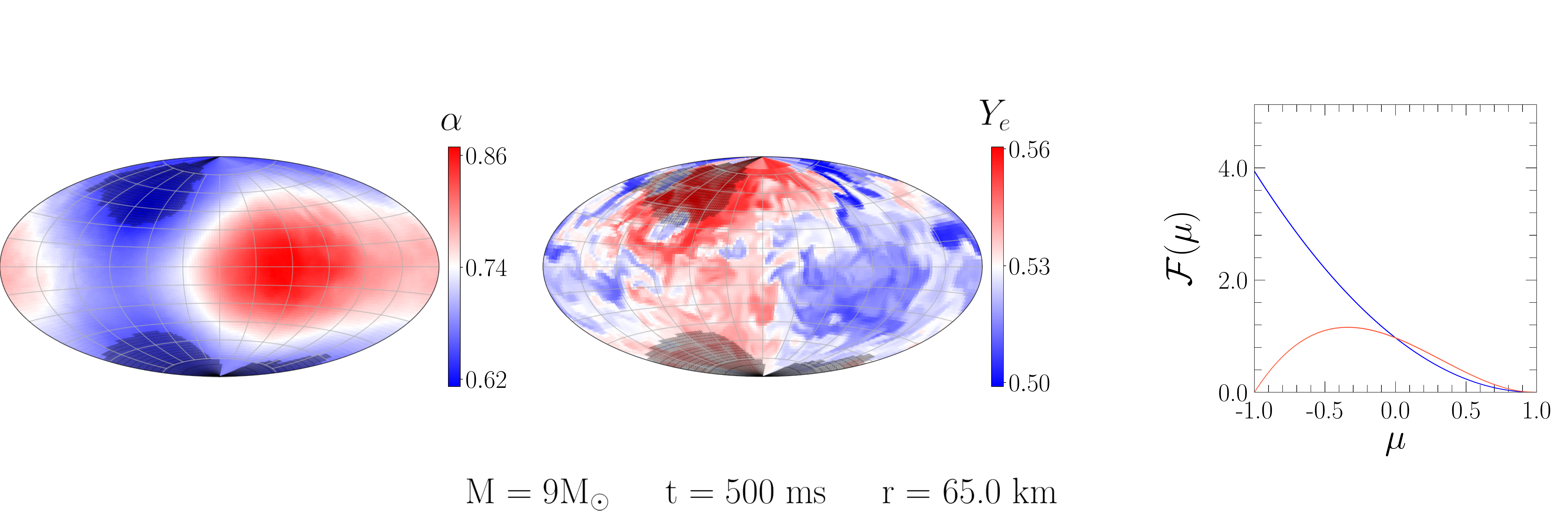}
       \caption{Left and middle panels: Aitoff projection of $\alpha$ and $Y_e$ where the shaded areas  represent the SN zones with ELN crossings in the range $[50,150]$~km. 
       Right panels: The shapes of $\mathcal{F}(\mu)$ that satisfy Eq.~(\ref{eq:crossing_condition}).
       All the panels are for the 9$\mathrm{M_{\odot}}$ model at $r=65$ km, and
       the upper and lower panels show the data from
         the $t = 400$ 
       and 500 ms  snapshots, respectively.  
       Note that the crossings occur at locations where $\alpha$ has nearly its minimal value,   
       in contrast  to what we found for forward crossings.
       The pattern of $\alpha$ (and $Y_e$) is a consequence of LESA~\cite{Glas:2018vcs}.}
        \label{Fig:9Msol_400_500ms_F_alpha_Ye}
\end{figure*}

\subsubsection*{$\mathbf{t=400}$  \rm{\textbf{ms}} }

\begin{figure}[tbh!]
  \centering
    \includegraphics[width=0.35\textwidth]{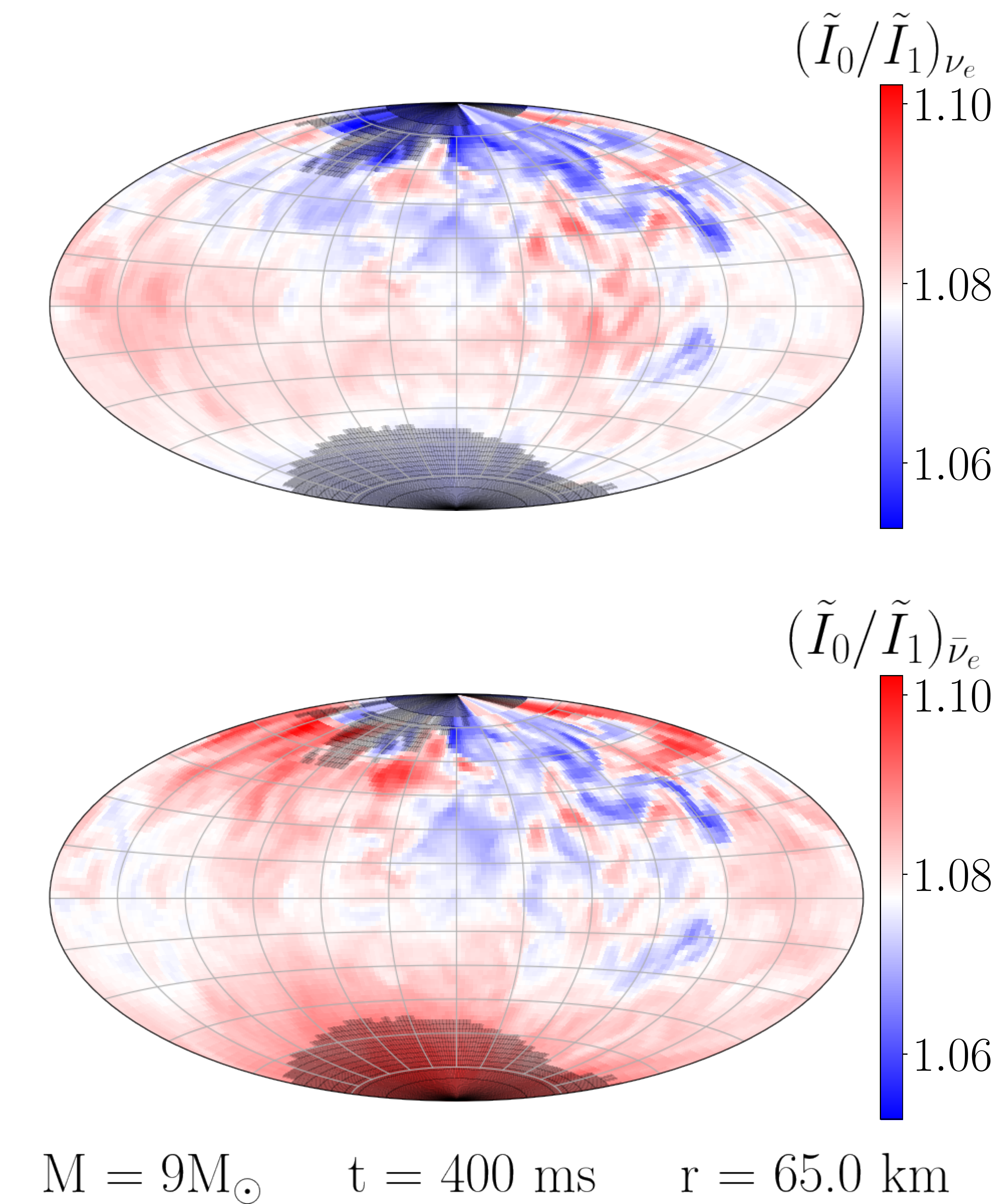}
       \caption{
       Aitoff projections of $\tilde{I}_0/\tilde{I}_1$ for $\nu_e$ (upper panel) and $\bar{\nu}_e$ (lower panel).
        Note that $(\tilde{I}_0/\tilde{I}_1)$ is larger  for $\bar{\nu}_e$ at the points 
        where the ELN crossings occur (indicated by shaded areas), suggesting  that $\bar{\nu}_e$ undergoes more scatterings there. 
        }
        \label{Fig:9Msol_400ms_I0_I1}
\end{figure}

At this time, the explosion has already set in with the shock radius being at $\sim1100$ km.
The  ELN crossings exist below the shock, from $r=56$ km (for which $\sqrt2 G_{\mathrm{F}} n_{\nu_e} = 5.1\times 10^4$ km$^{-1}$) up to $r=1500$ km, 
which is beyond the shock and where we stop our analysis.  It seems that in this case only  ``backward" crossings have been captured, as can be seen from
the upper right  panel  of Fig.~\ref{Fig:9Msol_400_500ms_F_alpha_Ye}.
  However, it should be noted that this type of backward crossings is very different from the one already found 
in the $t=100-300$ ms snapshots. These crossings occur at smaller radii
and more precisely, in the SN post-shock region.
Hence, they should be mainly attributed to the scattering on free nucleons rather than 
  heavy nuclei, 
which  are already disintegrated almost completely in the SN post-shock region.
In particular, these crossings  tend to occur in the SN zones where the
$\bar\nu_e$ ($\nu_e$) angular distribution is more (less) isotropic.
This is indicated in 
Fig.~\ref{Fig:9Msol_400ms_I0_I1}, which shows the Aitoff projections of $\tilde{I}_0/\tilde{I}_1$
 for  $\nu_e$ and $\bar{\nu}_e$, where 
 \begin{equation} 
  \tilde{I}_n = 
  \sqrt2 G_{\mathrm{F}}
 \int_{-1}^{1} \mathrm{d}\mu\ \mu^n \int_0^\infty \int_0^{2\pi} \frac{E_\nu^2 \mathrm{d} E_\nu \mathrm{d} \phi_\nu}{(2\pi)^3}
        f_{\nu}(\mathbf{p}).
 \label{Eq:G}
\end{equation}
This quantity is basically a measure of the isotropy of the neutrino angular distribution:
$\tilde{I}_0/\tilde{I}_1=1$  when the distribution is perfectly peaked in the forward direction, whereas $\tilde{I}_0/\tilde{I}_1=\infty$ when it is completely isotropic.  
Note that $\tilde{I}_n$ and $I_n$ are related by $I_n = (\tilde{I}_n)_{\nu_e} - (\tilde{I}_n)_{\bar\nu_e}$.
For the locations where these crossings are observed, antineutrinos have a larger $\tilde{I}_0/\tilde{I}_1$, 
meaning that they are slightly more isotropic. 

Somewhat surprisingly, 
 these  backward ELN crossings 
 show strong correlation with the elctron fraction,  $Y_e$, 
i.e. they occur (or at least they are easier to find) where $Y_e$ is maximal, 
as shown in the middle panels of Fig.~\ref{Fig:9Msol_400_500ms_F_alpha_Ye}. 
One can then also understand the correlation of these crossings with  $\alpha$, i.e. they occur where $\alpha$ is minimal:
In the region where $\alpha$ is minimal (due to LESA), $\nu_e$ absorption is more dominant, 
thus driving the ejecta to higher $Y_e$.
Although making strong conclusions  on the exact underlying physics of 
this phenomenon
is extremely difficult, 
 \textit{a plausible} explanation could be that 
  in the SN zones where $Y_e$ is maximal,
 $\bar\nu_e$ ($\nu_e$) is decoupled at
 larger (smaller) radii due to stronger (weaker) charged-current interactions.
 Moreover, the ratio of the  neutral-current cross sections of neutrinos and antineutrinos
  is slightly larger when the target particles are protons due to the weak-magnetism
corrections~\cite{Horowitz:2001xf}.
Both of these will result in a $\bar\nu_e$ ($\nu_e$) angular distribution 
which is more (less) isotropic in the SN zones where $Y_e$ is 
maximal. It is illuminating to note that there is indeed a competition between two effects here. 
While having
 smaller values of $\alpha$ reduces the chance of crossings~\cite{Shalgar:2019kzy, Nagakura:2019sig}, 
the more isotropic  neutrino angular distributions (of $\bar\nu_e$) enhance it.
Here, it is the 
latter that apparently wins the competition (though there is no quantitative way to describe
this competition).

The reason why this type of crossings is missing in previous time snapshots is that the explosion sets in only at about 300 ms. 
In non-exploding models and before the explosion sets in, the density gradient around the PNS is more shallow. Therefore the decoupling radii of $\nu_e$ and $\bar{\nu}_e$ are at larger values and also have a larger separation between each other. This leads to relatively large differences of the angular distributions of the neutrinos at given radius exterior to the decoupling region and thus to a reduced relative importance of nucleon scattering between the PNS and the shock. Therefore, nucleon scattering cannot overrule the stronger intrinsic backward component of the neutrino distribution.   
After the explosion has set in, the density gradient between the PNS and the shock steepens, because the density behind the shock declines, and the neutrinospheres of $\nu_e$ and $\bar{\nu}_e$ move closer together. In this case, at exterior radii, the $\nu_e$ and $\bar{\nu}_e$ angular distributions as emitted from their neutrinospheres are less dissimilar, and the nucleon scatterings and charged-current absorptions
can affect the ELN distribution relatively more strongly such that backward crossings can occur. 


\subsubsection*{$\mathbf{t=500}$  \rm{\textbf{ms}} }
In this snapshot, we find ELN crossings  starting from $r=68$  km (for which $\sqrt2 G_{\mathrm{F}} n_{\nu_e} = 2.6\times 10^4$ km$^{-1}$) up to $r=1500$ km, 
where we stop our analysis. 
As indicated in the lower  panels of Fig.~\ref{Fig:9Msol_400_500ms_F_alpha_Ye}, we again can only  capture  ``backward" crossings which
show  anti-correlations with  $\alpha$ distribution  below the shock,  analogously to what happens at $t=400$ ms.

\subsection{$20\mathrm{M_{\odot}}$ supernova model}
\subsubsection*{$\mathbf{t=100}$  \rm{\textbf{ms}} }
In this time snapshot, we find ELN crossings  starting from $r=105$ km (for which $\sqrt2 G_{\mathrm{F}} n_{\nu_e} = 1.1\times 10^5$ km$^{-1}$) up to $r=1500$ km, 
where we stop our analysis. 
While at lower radial distances we capture forward  crossings,  at larger radii (above the shock which is at 
$\sim150$ km),
 the captured crossings are in the backward direction as expected.

\begin{figure}
  \centering
    \includegraphics[width=.4\textwidth]{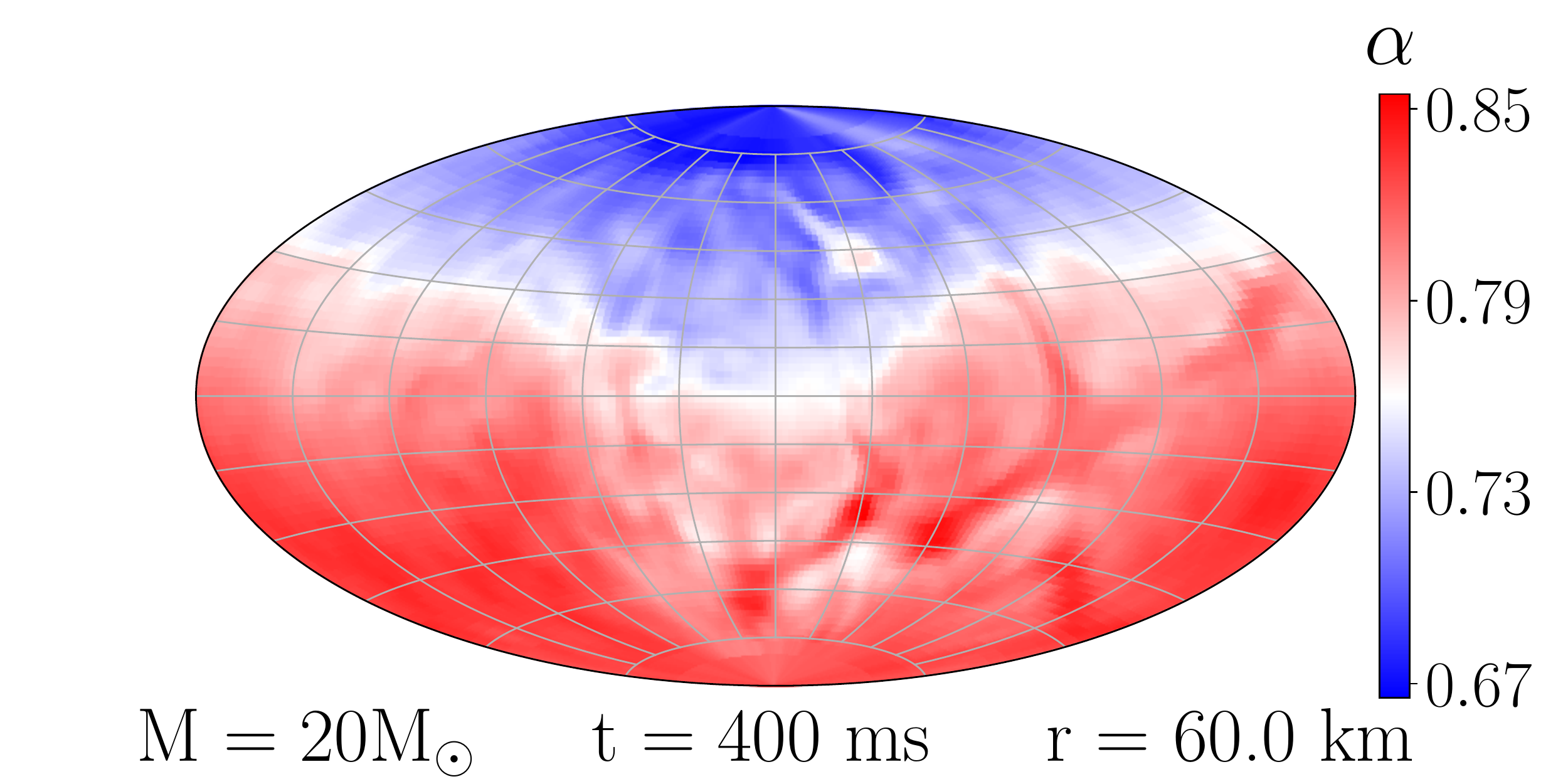}
       \caption{Aitoff projection of $\alpha$ for the 20$\mathrm{M_{\odot}}$ model in the  $t=400$ ms snapshot at $r=60$ km. 
      We observe that  the average value of $\alpha$ is larger than the one in the 9$\mathrm{M_{\odot}}$ model.}
        \label{Fig:20Msol_alpha}
\end{figure}

\begin{figure}[tbh!]
  \centering
    \includegraphics[width=.35\textwidth]{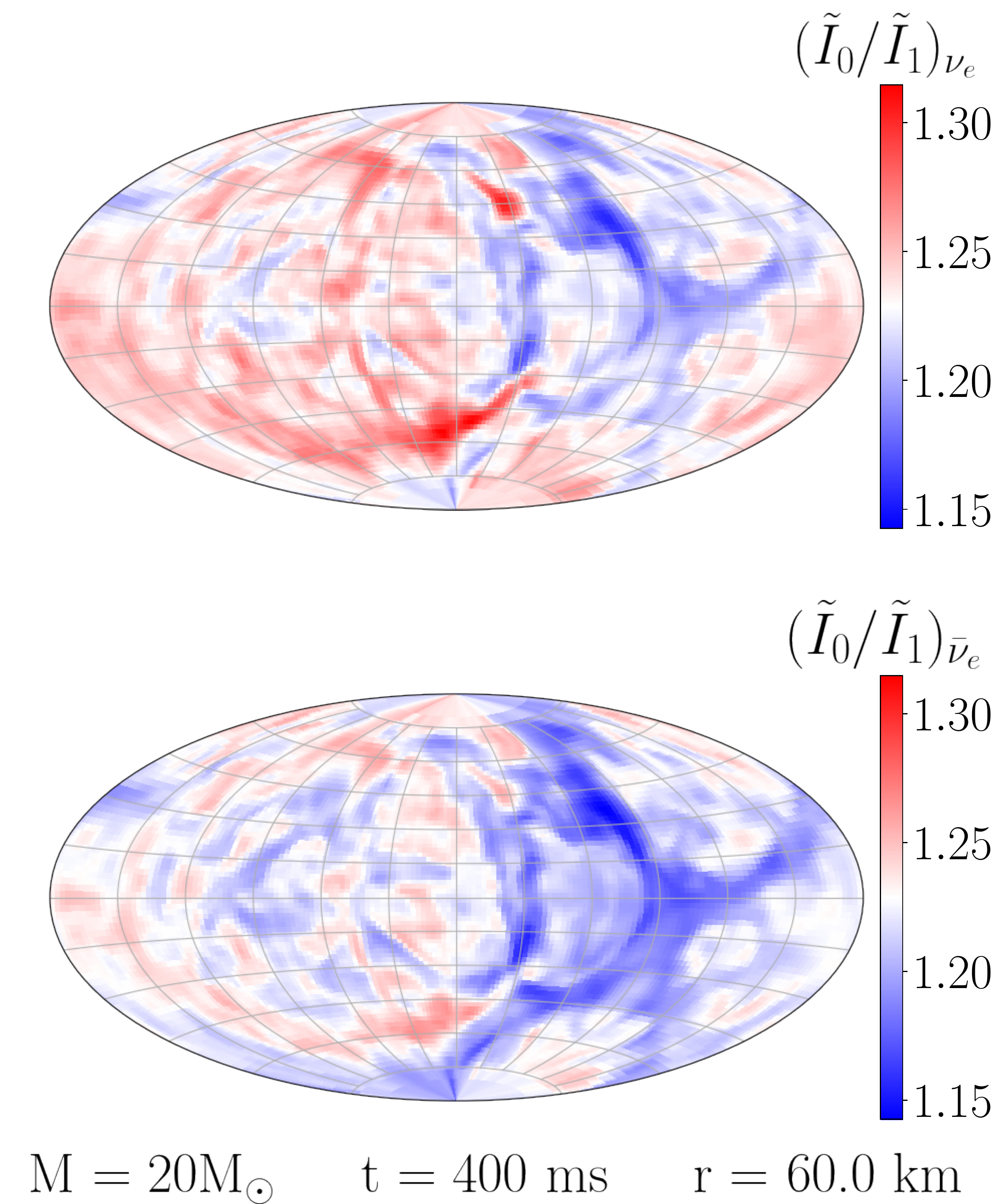}
       \caption{Aitoff projections of $\tilde{I}_0/\tilde{I}_1$ for the 20$\mathrm{M_{\odot}}$ model in the  $t=400$ ms snapshot at  $r=60$ km, for $\nu_e$ (upper panel) and $\bar\nu_e$ (lower panel). 
       }
        \label{Fig:20Msol_I0_I1}
\end{figure}

\subsubsection*{$\mathbf{t=200-500}$  \rm{\textbf{ms}} }
In these time snapshots, 
we find no ELN crossing  below the shock, though
the usual backward ubiquitous ELN crossings 
   above the shock (starting from  $\simeq 300$ km) can be observed. 
This is   different from what we observed in the 9$\mathrm{M_{\odot}}$ model. 
The failure of explosion has indeed  important consequences for the occurrence of ELN crossings.
As a matter of fact, $\tilde{I}_0/\tilde{I}_1$ is relatively large for both neutrinos and antineutrinos
in the  post-shock layer of the 20$\mathrm{M_{\odot}}$ model, as shown in Fig.~\ref{Fig:20Msol_I0_I1}. 
This comes from the fact that  the shock in 
this non-exploding  model retracts after $\sim100$ ms post bounce. 
Therefore, on the one hand the post-shock region is (on average) closer to the neutrinospheres, where the neutrino distributions are more isotropic. 
On the other hand, 
neutrinos and antineutrinos experience more scatterings which 
means that their angular distributions  become more  similar,
reducing the chance for the occurrence of ELN crossings.
This demonstrates  that a large value of $\alpha$ alone (Fig.~\ref{Fig:20Msol_alpha}) 
 is not sufficient for the occurrence of forward crossings, but $\tilde{I}_0/\tilde{I}_1$ also contains crucial information. 
The high matter density also suppresses backward crossings in the post-shock region 
of this model as discussed previously,
i.e. the intrinsic population of $\nu_e$ in the backward direction is so large that
scattering-induced backward traveling $\bar\nu_e$ cannot generate an ELN crossing.

\section{Linear stability analysis}

Until now, we have only focused on the occurrence of ELN crossings in the SN environment.
 Here we consider flavor instabilities 
associated with the backward ELN crossings and we discuss some of their important 
characteristics.

The flavor content of a neutrino gas at each  space-time point and 
 in the momentum mode $\mathbf{p}$, can be determined  by its flavor
density matrix $\varrho_{\mathbf{p}}(t, \mathbf{r})$~\cite{Sigl:1992fn} 
\begin{align}
  \varrho = \frac{f_{\nu_e} + f_{\nu_x}}{2}
  + \frac{f_{\nu_e} - f_{\nu_x}}{2} \begin{bmatrix}
    s & S \\ S^* & -s \end{bmatrix},
\end{align}
where 
the complex and real scalar fields $S$ and
$s$ describe the flavor coherence and flavor conversion of 
neutrinos, respectively. One can  similarly define the flavor
density matrix, $\bar\varrho$, of  antineutrinos.
In the linear regime, where significant flavor conversions have not occurred yet
($|S_\mathbf{v}|\ll1$ and $s\approx 1$),
the evolution of the flavor coherence term 
 is governed by
\cite{Banerjee:2011fj, Vaananen:2013qja, Izaguirre:2016gsx}
\begin{equation}
 i (\partial_t + \mathbf{v} \cdot  \boldsymbol{\nabla} ) S_\mathbf{v}
 = (\epsilon_0 + \mathbf{v} \cdot
 \boldsymbol{\epsilon} ) S_\mathbf{v}
 - \int\!\mathrm{d}\Gamma_{\mathbf{v}'} (1 - \mathbf{v} \cdot \mathbf{v}')
 \G_{\mathbf{v}'} S_{\mathbf{v}'},
 \label{Eq:linear}
 \end{equation}
 where $\mathrm{d}\Gamma_{\mathbf{v}'}$ is the differential solid angle in the
 direction of $\mathbf{v}'$,
 $\epsilon_0 = \lambda+\int\!\mathrm{d}\Gamma_{\mathbf{v}'} \G_{\mathbf{v}'}$,
 and $\boldsymbol{\epsilon} = \int\!\mathrm{d}\Gamma_{\mathbf{v}'}
 \G_{\mathbf{v}'} \mathbf{v}'$. Here we have ignored the vacuum term in the Hamiltonian, 

Given the fact that Eq.~(\ref{Eq:linear}) is a linear equation of motion, it has solutions of the form
 $ S_\mathbf{v}(t,\mathbf{r}) = Q_\mathbf{v}\,  e^{-i\Omega t +
    i\mathbf{K}\cdot\mathbf{r}}$,
where $\Omega$ and $\mathbf{K}$ are independent of $\mathbf{v}$.
The neutrino gas is then unstable in the flavor space
 ($S_\mathbf{v}$ can grow exponentially) if Eq.~(\ref{Eq:linear}) features
 temporal instabilities, i.e.,
 for some real 
$\mathbf{K}$ there exists a complex $\Omega$
which has  a positive imaginary component, namely
$\Omega_\mathrm{i}=\mathrm{Im}(\Omega) >0$. 
Such unstable modes with fast growing amplitudes can exist when
 $G_\mathbf{v}$ crosses 0 at some angle(s)
~\cite{Izaguirre:2016gsx,Dasgupta:2016dbv,Capozzi:2017gqd, Abbar:2017pkh,Yi:2019hrp,Capozzi:2019lso}.

\begin{figure}
  \centering
    \includegraphics[width=.45\textwidth]{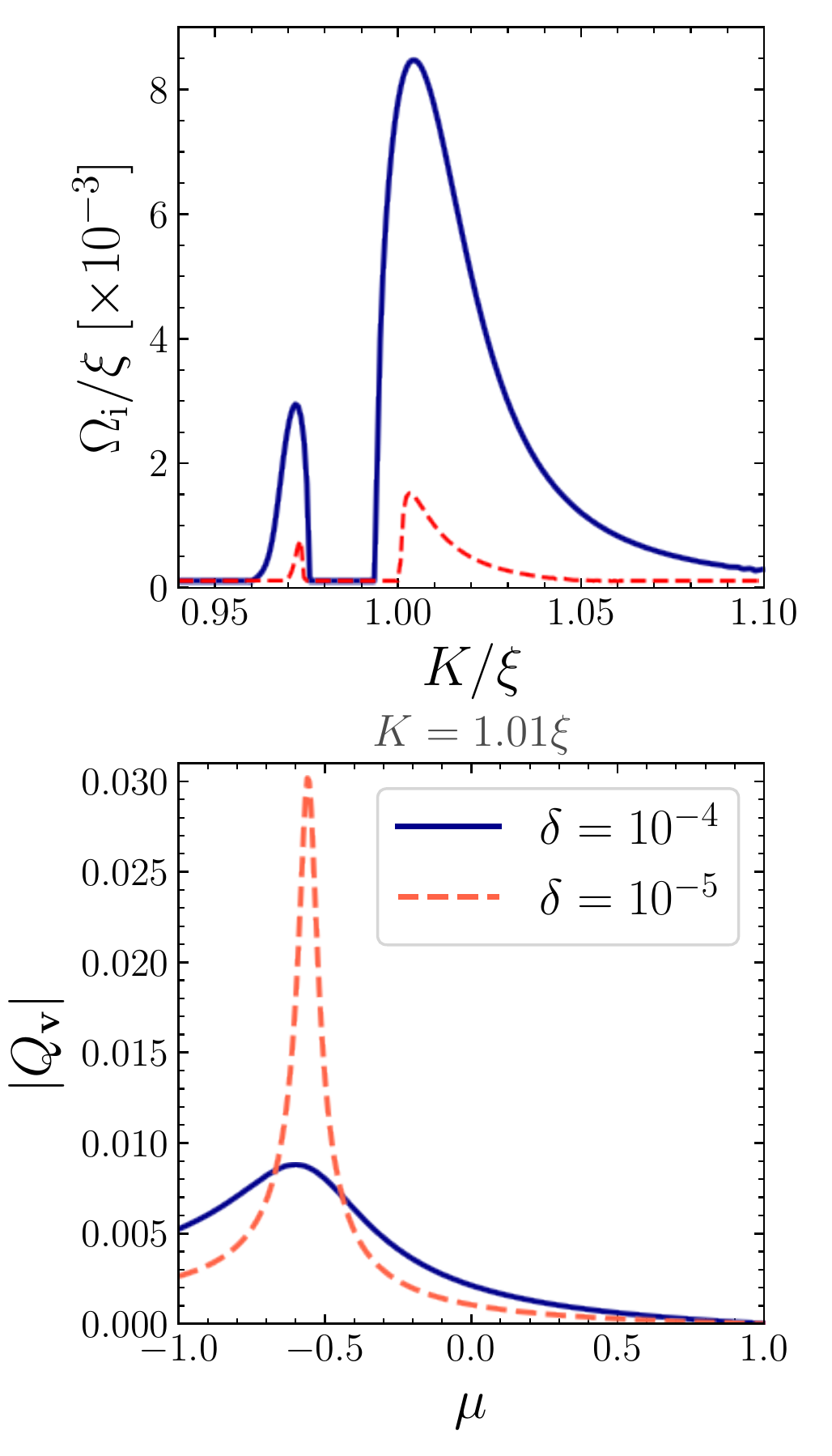}
       \caption{Upper panel: Growth rate of temporal fast instabilities 
       as functions of the real wave number $K$ of the fast  modes propagating in the radial direction.
       Note that despite the fact that
       the crossings in the backward direction are very shallow, the growth rates can still be significant for
       large neutrino number densities.
       For instance, for $\xi = 10^4$ km$^{-1}$ and $\delta=10^{-4}$, the growth rate can be as high as $~10^2$ km$^{-1}$. 
       Lower panel: An example of the angular distributions of the  eigenvectors 
of unstable modes, $Q_\mathbf{v}$, which is normalized to one here. 
These distributions correspond to the unstable modes at  $K=1.01\xi$, where
the growth rates  are close to their maxima.
Note that the amplitude of $Q_\mathbf{v}$ is extremely small in the forward
direction where the majority of neutrinos are traveling.
In these calculations we assumed that the 
neutrino gas possesses axial symmetry.
       }
        \label{fig:Qkappa}
\end{figure}

Although  ELN crossings in the backward direction can be  extremely narrow, their corresponding
fast growth rates are not necessarily small. This is illustrated in the upper panel of 
Fig.~\ref{fig:Qkappa}, where the growth
rate of temporal fast instabilities, $\Omega_\mathrm{i}$, is shown for an schematic ELN distribution, defined as
 \begin{equation}\label{eq:schem}
 G(\mu) = \xi \times \left\{
                \begin{array}{ll}
                  +1\quad \text{for}\quad \mu >\mu_c,\\
                   -\delta \quad \text{for}\quad \mu<\mu_c.
                \end{array}
              \right.
\end{equation}
Here, $\mu_c$ is the point where the  crossing occurs, $\delta$ is the depth of the backward 
scattering, and $\xi$ determines the overall scale of the ELN distribution. 
In our calculation we considered  $\mu_c = 0.95$ and $\delta= 10^{-4}$ and $10^{-5}$, which is appropriate for
the backward ELN crossings at large radii.
 The presented growth rates correspond to  the temporal unstable modes  
 propagating in the radial direction,
assuming that the axial symmetry is preserved in the neutrino gas. 

Despite the possibility of  having significant growth rates,
one should be very careful when interpreting  
 the fast instabilities corresponding
to the backward ELN crossings. An extremely   important issue,
 which has been completely overlooked  in the literature,
is associated with the shape of the angular distribution of the  eigenvectors 
of unstable modes. As indicated in the lower panel of
Fig.~\ref{fig:Qkappa}, the eigenvectors of the unstable fast modes are extremely
peaked in the backward direction, where the population of neutrinos is extremely small.
This comes as no surprise since the ELN crossings are also generated by this
small population. 
However, this means that in the linear regime, the unstable
modes can mostly affect  the neutrinos propagating in the backward 
direction, which are a tiny fraction of all neutrinos. 
Despite this,  significant flavor conversions can still occur in the nonlinear regime.
Hence, to have a better understanding of this issue,  sophisticated  neutrino flavor conversion simulations
in the nonlinear regime are absolutely indispensable.

\section{Discussion and Conclusions}
To assess  the occurrence of ELN crossings and their   associated
 fast neutrino flavor instabilities  in CCSNe,
we have analyzed two recent state-of-the-art 3D CCSN simulations of a 9$\mathrm{M_{\odot}}$ and a 20$\mathrm{M_{\odot}}$
progenitor model,
using the method proposed recently in Ref.~\cite{Abbar:2020fcl}.  
 Not only do we  confirm the results previously reported in the literature using post-processing 
 calculations~\cite{Abbar:2018shq} 
 or 1D/2D models~\cite{Morinaga:2019wsv,Nagakura:2019sig} , 
but  we also highlight some new insights. Firstly, 
we bring up the  possibility of 
 having backward ELN crossings in the SN post-shock region. 
 Secondly, we argue that the failure  of the SN explosion
can have important consequences for the existence of ELN  crossings, i.e.
it can seriously hinder their occurrence in the SN environment.   

In the case of  the successfully exploding 9$\mathrm{M_{\odot}}$ SN model, we observed  
a large number of ELN crossings in the  forward direction in the 100 ms and 200 ms snapshots
within/above the neutrino decoupling region (see Fig.~\ref{Fig:9Msol_100ms_alpha}). 
The spatial distribution of these ELN crossings is correlated with the distribution of  $\alpha$,
which itself originates from  the  convective activity inside the PNS  
(which is thought to generate LESA at later times, namely $t_{\mathrm{pb}} \gtrsim 300$~ms in this model). 
However, such ELN crossings are not observed in the later snapshots. 
This may look strange at first, considering the correlation of these crossings with LESA
 which  becomes stronger at later times in this model.
 This means that one should expect 
a smaller difference between $n_{\nu_e}$ and $n_{\bar{\nu}_e}$ 
(in the hemisphere opposite to the LESA dipole direction) 
and consequently, a higher chance for the occurrence of forward ELN crossings.
Despite this, 
as time goes by the difference between the radii 
of the neutrinospheres of $\nu_e$ and $\bar{\nu}_e$ decreases, 
making their angular distributions more similar in  the forward direction.
As a result, 
the range of $\mu$ where $G(\mu$) changes its sign becomes extremely narrow. 
In the context of our method, this means that although forward ELN crossings might be more likely to occur at later times~\cite{Abbar:2018shq,Nagakura:2019sig}, 
   higher neutrino angular moments are required to capture them.

We also observe a ubiquitous distribution of  ELN crossings  in the backward direction in the pre-shock SN region
in all time snapshots of the 9$\mathrm{M_{\odot}}$ model 
as well as the 20$\mathrm{M_{\odot}}$ one.
Such ELN crossings, first reported in Ref.~\cite{Morinaga:2019wsv} in a 1D SN model, are thought to be generated 
by coherent neutrino scattering on heavy nuclei, which are
enhanced for the slightly more energetic $\bar{\nu}_e$. 
Our results confirm the presence 
of these crossings in multi-D SN models and show that they are probably very  generic. 

Somewhat surprisingly,  we also find backward ELN crossings 
 in the post-shock region of the  400 and 500 ms  snapshots
of the 9$\mathrm{M_{\odot}}$  model
  at radii which can be as small as $\sim50$ km. 
  Unlike the backward ELN crossings in the SN pre-shock region,
 which are generated by  coherent scatterings on heavy nuclei, 
  the backward  crossings below the SN shock should be attributed to the scatterings on free nucleons since  
heavy nuclei here are  disintegrated almost completely. 

 A remarkable  feature of the backward ELN crossings 
in the SN post-shock  region
 is that their spatial distribution
shows a strong correlation  with  $Y_e$
 (which leads to an anti-correlation with $\alpha$),
 in contrast to the forward crossings. 
  Though it is very difficult to draw any conclusion about the exact nature of this pattern,
  it could be explained, at least partially,  by noting that
   in these SN zones where $Y_e$ is larger, 
 $\bar\nu_e$ ($\nu_e$) decouples at
 larger (smaller) radii due to stronger (weaker) charged-current interactions.
 Moreover, the neutral-current scattering of $\bar\nu_e$ on free nucleons is
 slightly enhanced therein due to the larger fraction of protons.
The presence of such backward ELN crossings implies that at least at later times, one could expect
an almost ubiquitous distribution of ELN crossings below the SN shock,  forward crossings
around the minima of  $Y_e$ and  backward crossings around its maxima.

As for the non-exploding 20$\mathrm{M_{\odot}}$  model,
 in the post-shock region we observe forward crossings   only in the 100 ms snapshot, 
 whereas backward crossings are ubiquitous in the pre-shock region. 
As a matter of fact, the failure  of the explosion  has significant consequences for
the occurrence of fast instabilities.
Due to the failure of the  explosion, the post-shock region is closer to the neutrinospheres
where the neutrino angular distributions are more isotropic. In addition,   
the matter density below the shock  
increases with time as the shock retracts, which 
implies more scatterings for  $\nu_e$  and $\bar\nu_e$. This means that their angular distributions
 become more isotropic  (and therefore more similar) which in turn  
reduces the chance for the occurrence of forward ELN crossings. Note that this happens
despite the fact that the overall value of $\alpha$ is larger in this case compared to the
9$\mathrm{M_{\odot}}$ SN model. 
This clearly shows that one needs care when assessing the occurrence of forward ELN crossings
 by just considering the parameter $\alpha$\footnote{Note that this case is different
 from the case studied in Ref.~\cite{Abbar:2019zoq} where the overall value of $\alpha$ was very small
 for the non-exploding model.}.

A failed  explosion can similarly explain why there are no backward ELN crossings
in the post-shock layer of the 20$\mathrm{M_{\odot}}$ SN model. 
Due to the high matter density below the shock and the closer distance to the PNS, the backward direction
is still dominated  by preexisting  $\nu_e$'s, and the higher energetic  $\bar\nu_e$'s
cannot generate an ELN crossing in the backward direction.
This is also why the backward ELN crossings in the post-shock region
 are not observed in the earlier 
snapshots of the 9$\mathrm{M_{\odot}}$ model where the shock is still relatively
close to the PNS and the matter density is high.

Note that this argument only suggests that the large matter densities behind the stalled or retreating shock
can \textit{hinder} the occurrence of ELN crossings. 
However, one can not make any quantitative estimate on the 
exact level of suppression of the ELN crossings, 
which depends on a number of factors such as the SN model, its mass, the fate of the explosion, the neutrino transport method, etc.
This means that though less probable, ELN crossings can still occur in the post-shock region of non-exploding models.
As an example, in the context of 1D
models,  Ref.~[61] found a number of backward ELN crossings in the post-shock  region of a non-exploding  
20$\mathrm{M_{\odot}}$  model in the presence of
muon creation in the SN core, which is not included in the models analyzed herein.
Overall, this  highlights the importance of  more realistic modelings of CCSNe to have 
a better understanding 
of the occurrence of ELN crossings and, therefore, of neutrino flavor conversions
therein, 
given the fact that the fate of the explosion may remarkably affect the reliability of our predictions of this
phenomenon.

In summary, apart from confirming the previous reports of 
the occurrence of ELN crossings and fast instabilities  by the analysis of our
3D state-of-the-art CCSN simulations, we point out a new possibility for 
the occurrence of probably scattering-induced ELN crossings
in the post-shock region of exploding SNe.
Nevertheless, it should be kept in mind that
the occurrence of fast instabilities does not necessarily
lead to (significant) flavor conversions~\cite{Padilla-Gay:2020uxa}.  
Our results also  suggest that the occurrence of ELN crossings
 could be strongly hindered in non-exploding SN models with high mass accretion rates.   
A  key limitation of our study is that it is  based on only a few of the neutrino angular moments and therefore, 
our understandings are still somewhat speculative.      
In particular, drawing strong conclusions  from a thus found crossing 
on the underlying physics is extremely difficult. 
In order to have a more accurate picture of the physics of fast neutrino flavor conversions
 in CCSNe,
one needs to consider the full phase-space distribution functions of neutrinos.  
Therefore, 3D CCSN simulations with Boltzmann neutrino transport are indispensable. 
Moreover, the occurrence of ELN crossings should be studied for a wide range of  progenitor masses, which can offer accretion and explosion conditions that are considerably different from those of the considered 9$\mathrm{M_{\odot}}$ and 20$\mathrm{M_{\odot}}$ cases because of the progenitor-specific accretion histories after core bounce.

\section*{Acknowledgments}
We would like to thank Georg Raffelt for his comments on the manuscript and useful discussions.  
We   acknowledge partial support by the Deutsche Forschungsgemeinschaft (DFG, German Research Foundation)
through Grant 
Sonderforschungsbereich (Collaborative Research Center) SFB1258 ``Neutrinos and Dark Matter in Astro- and Particle Physics (NDM)''.
Additionally, F.C.'s work at Virginia Tech is supported by the U.S. Department of Energy under the award number DE-SC0020250 and DE-SC0020262 and
at Garching, funding by the
European Research Council through Grant ERC-AdG No.~341157-COCO2CASA
and by DFG under Germany's Excellence Strategy through
Cluster of Excellence ORIGINS (EXC-2094)---390783311 is acknowledged.
Computer  resources  for  this  project  have been  provided  by  the  Leibniz  Supercomputing  
Centre (LRZ)  under  LRZ  project  ID  pr62za  and  by  the  Max Planck Computing and Data Facility (MPCDF) on the HPC system Hydra.
I.T. acknowledges support from  the Villum Foundation (Project No.~13164), the Danmarks Frie Forskningsfonds (Project No.~8049-00038B) and the Knud H\o jgaard Foundation.

\bibliographystyle{JHEP}
\bibliography{Biblio}

\end{document}